\documentclass[cits]{PoS}

\usepackage[utf8]{inputenc} 
\usepackage{amsmath} 
\usepackage{slashed}

\newcommand{\triph}{\lambda_{HHH}^{}}

\title{Impact of heavy sterile neutrinos on the triple Higgs coupling}

\ShortTitle{Neutrinos and the triple Higgs coupling}

\author{Julien Baglio\\
       Institute for Theoretical Physics,  University of
  T\"ubingen, Auf der Morgenstelle 14, 72076~T\"ubingen, Germany\\
Institute for Advanced Study, Durham University, Cosin's Hall, Palace
Green, Durham DH1 3RL, United Kingdom\\
Institute for Particle Physics Phenomenology, Department
  of Physics, Durham University,  South Road, Durham DH1 3LE, United
  Kingdom\\
       E-mail: \email{julien.baglio@uni-tuebingen.de}}

\author{\speaker{Cédric Weiland}\\
        Institute for Particle Physics Phenomenology, Department
  of Physics, Durham University,  South Road, Durham DH1 3LE, United
  Kingdom\\
        E-mail: \email{cedric.weiland@durham.ac.uk}}

\abstract{New physics beyond the Standard Model is required to give
  mass to the light neutrinos. One of the simplest ideas is to
  introduce new heavy, gauge singlet fermions that play the role of
  right-handed neutrinos in a seesaw mechanism. They could have large
  Yukawa couplings to the Higgs boson, affecting the Higgs couplings
  and in particular the triple Higgs coupling $\triph$, the measure of
  which is one of the major goals of the LHC and of future
  colliders. We present a study of the impact of these heavy neutrinos
  on $\triph$ at the one-loop level, first in a simplified 3+1 model
  with one heavy Dirac neutrino and then in the inverse seesaw model.
  Taking into account all possible experimental
  constraints, we find that sizeable deviations of the order of 35\% are possible, 
  large enough to be detected at future colliders, making the triple Higgs coupling a
  new, viable observable to constrain neutrino mass models. The
  effects are generic and are expected in any new physics model
  including TeV-scale fermions with large Yukawa couplings to the
  Higgs boson, such as those using the neutrino portal. {\footnote{Preprint: IPPP/17/77}} }

\FullConference{The European Physical Society Conference on High Energy Physics\\
		5-12 July, 2017\\
		Venice}

\begin{document}

\section{Introduction}

The Super-Kamiokande experiment firmly established in 1998 that neutrinos
oscillate~\cite{Fukuda:1998mi}, which calls for an extension of the Standard Model (SM) 
that generates neutrino masses and mixing. One of the simplest possibilities to explain neutrino masses
is to add new fermionic gauge singlets that play the role of
right-handed neutrinos. These new fermionic
states could have large Yukawa couplings to the Higgs boson, having a sizeable impact on the Higgs couplings and opening new search strategies.

The Higgs self-couplings, and in particular the triple Higgs coupling 
$\triph$, play a central role in probing electroweak symmetry
breaking (EWSB) induced by the Higgs mechanism~\cite{Higgs:1964ia,Englert:1964et,Higgs:1964pj,Guralnik:1964eu,Higgs:1966ev}. 
The measure of $\triph$ is one of the major goals of the LHC and of
the future planned colliders such as the electron-positron
International Linear Collider (ILC) or the Future Circular Collider in
hadron mode (FCC-hh), a potential 100 TeV follow-up of the
LHC. Investigating possible beyond-the-SM (BSM) effects on this coupling is thus very much
needed and the effects induced by the heavy neutrinos present in seesaw
mechanisms have been overlooked so far.

We present a study of the impact of these heavy neutrinos on $\triph$,
first by considering a simplified 3+1 model where the SM is minimally
modified to account for three light massive Dirac neutrinos and one
heavy sterile Dirac neutrino; then by considering the inverse seesaw
mechanism~\cite{Mohapatra:1986bd,Mohapatra:1986aw,Bernabeu:1987gr}
which is a realistic, renormalisable mass model with 9 Majorana
neutrinos. Taking into account all theoretical and experimental
constraints, we find in both
studies~\cite{Baglio:2016ijw,Baglio:2016bop} sizeable effects, of the
order of 35\% for large off-shell Higgs momentum $q_H^*$ and of the
order of 10\% for $q_H^* = 500$~GeV. This is clearly detectable at the
FCC-hh and may be probed at the ILC, making the triple
Higgs coupling $\triph$ a new, viable observable for the neutrino sector
in order to constraint mass models.

\section{The triple Higgs coupling}

The Higgs field $\Phi$ of the SM can be written as
\begin{align}
 \Phi = & \frac{1}{\sqrt{2}}
          \left(\begin{matrix} \sqrt{2} G^+\\ \mathrm{v}+H+\imath
                  G^0\end{matrix}\right)\,,
\end{align}
where $H$ is the Higgs boson, $G^0$ is the neutral Goldstone boson,
$G^\pm$ are the charged Goldstone bosons, and $\mathrm{v}\simeq
246$~GeV is the Higgs vacuum expectation value. After EWSB, the scalar
potential of the SM contains the following terms involving the Higgs
boson $H$,
 \begin{align}
  V(H)= \frac{1}{2} M_H^2 H^2 + \frac{1}{3!} \lambda_{HHH} H^3 +  \frac{1}{4!} \lambda_{HHHH} H^4\,,
 \end{align}
where $M_H^{}$ is the Higgs boson mass and the tree-level values for
the triple and quartic Higgs couplings are $\lambda^{0}_{HHH}=-3
M_H^2/\mathrm{v}$ and $\lambda^{0}_{HHHH}=- 3 M_H^2/\mathrm{v^2}$
respectively.

Our one-loop calculation is performed in the on-shell renormalisation
scheme. Our Higgs and electroweak inputs are the Higgs mass $M_H^{}$,
the W and Z boson masses $M_W^{}$ and $M_Z^{}$, and the electric
charge $e$.
Details of the calculation and
analytical formulas can be found in our
articles~\cite{Baglio:2016ijw,Baglio:2016bop}. Our results will be
presented in terms of deviations with respect to the tree-level value
$\lambda^{0}_{HHH}$ and to the renormalised one-loop value in the SM
$\lambda_{HHH}^{1r,{\rm SM}}$ of the triple Higgs coupling,
\begin{align}
  \Delta^{(1)} \lambda_{HHH} & =
                               \frac{1}{\lambda^{0}}\left(\lambda_{HHH}^{1r,{\rm full}}
                               -\lambda^0\right)\,,\,\nonumber\\
  \Delta^{\rm BSM} & = \frac{1}{\lambda_{HHH}^{1r,{\rm
                     SM}}}\left(\lambda_{HHH}^{1r,{\rm full}}
                     -\lambda_{HHH}^{1r,{\rm SM}}\right)\,.
                     \label{eq:definedelta}
\end{align}
with $\lambda_{HHH}^{1r,{\rm full}}$ being the one-loop renormalised
triple Higgs coupling in the model considered.
We will compare our
results with the experimental sensitivities to the SM triple Higgs
coupling at the LHC, the ILC and the FCC-hh. We use a sensitivity of $\sim 35\%$ at
the high-luminosity run of the LHC (HL-LHC) according to
Ref.~\cite{CMS:2015nat} (see also Ref.~\cite{Campana:2016cqm}), with a
scaling of $1/\sqrt{2}$ to combine ATLAS and CMS results. Using
Ref.~\cite{He:2015spf} and again a rescaling we take a sensitivity of
$\sim 5\%$ at the FCC-hh with 3 ab$^{-1}$, and finally we take a
sensitivity of $10\%$~\cite{Fujii:2015jha} at the 1 TeV ILC with 5~ab$_{}^{-1}$.

\section{Simplified 3+1 model}
In a first study~\cite{Baglio:2016ijw}, we considered a simplified
model that includes 3 light neutrinos and an extra heavy neutrino. All
of them are Dirac fermions and the heavy neutrino couples to the SM
particles through its mixing with SM fields. This leads to the
following couplings between neutrinos and SM bosons, defined in the
mass basis,
\begin{align}
\mathcal{L}\ni & - \left(\frac{g_2^{}}{\sqrt{2}}\bar \ell_i^{} \slashed{W}_{}^{-}
                  B_{i j}^{} P_L^{} n_j^{}  + \frac{g_2^{}}{\sqrt{2}
                 M_W^{}} \bar \ell_i^{} G_{}^- B_{i j}^{}
                 (m_{\ell_i}^{} P_L^{}  - m_{n_j}^{} P_R^{} ) n_j^{}\right) +
                 \mathrm{H.c.}\nonumber \\
   & -\frac{g_2^{}}{2\cos \theta_W^{}} \bar n_i^{} \slashed Z
     (B^\dagger_{} B)_{i j}^{} P_L^{} n_j^{} + \frac{ \imath g_2^{}}{2
     M_W^{}} \bar n_i^{} (B^\dagger_{} B)_{i j}^{} G_{}^0 (-
     m_{n_i}^{} P_L^{} + m_{n_j}^{} P_R^{}) n_j^{}\nonumber \\
   & - \frac{g_2^{}}{2 M_W^{}} \bar n_i^{} (B^\dagger_{} B)_{i j}^{} H
     (m_{n_i}^{} P_L^{} + m_{n_j}^{} P_R^{}) n_j^{}\,,
\end{align}
where $\ell_i^{}$ are the charged leptons, $n_i^{}$ are the Dirac
neutrinos of mass $m_{1\cdots 4}^{}$, $g_2^{}$ is the $\mathrm{SU}(2)$
coupling constant, and $B$ is a $3\times4$ mixing matrix.

The most relevant experimental constraints on our model come from
electroweak precision observables (EWPO) and in particular from the
global fit performed in~\cite{delAguila:2008pw,deBlas:2013gla}. We
have also taken into account constraints on the mixing matrix $B$
coming from neutrino oscillations~\cite{Gonzalez-Garcia:2014bfa} with 
\mbox{$\delta_{CP}^{}=0$}. We also require two theoretical constraints. The
loop expansion has to remain perturbative, and we apply either a loose
(tight) bound of
\begin{equation}
 \left(\frac{\mathrm{max}|C_{i4}^{}|\,g_2^{}\,  m_{n_4}^{}}{2 M_W^{}}
 \right)_{}^3 < 16 \pi\,(2\pi)\,.
\end{equation}
Since for fixed mixing the heavy neutrino width grows with its mass,
we require as well \linebreak
\mbox{$\Gamma_{n_4}^{}\leq0.6\,m_{n_4}^{}$} for the quantum state to
be a definite particle.

For our numerical study, SM parameter values were taken from the
Particle Data Group~\cite{Patrignani:2016xqp} (with the exception of
the SM Higgs boson mass fixed to $M_H=125$~GeV).
\begin{figure}[!t]
 \centering
 \includegraphics[width=0.6\textwidth]{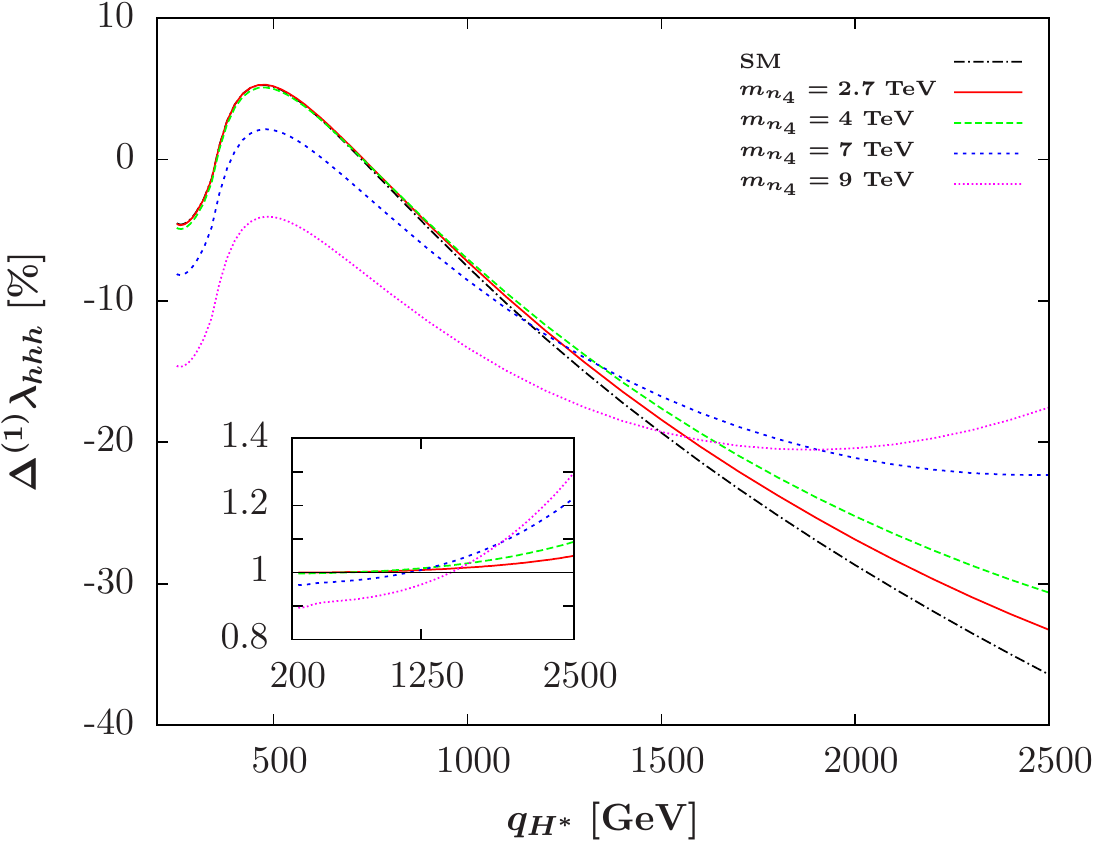}
\caption[]{One-loop corrections to the triple Higgs coupling
  $\triph$ (in \%) as a function of the momentum $q_{H^*}^{}$
  of the splitting $H^*(q_{H^*}^{})\to HH$ (in GeV). The ratio of the
  genuine BSM contribution to $\triph$ with respect
  to the one-loop SM contribution is shown in the insert.}
\label{hhh_dirac}
\end{figure}
Taking $B_{\tau4}^{}=0.087$, $B_{e4}^{}=B_{\mu4}^{}=0$,
Fig.~\ref{hhh_dirac} displays the one-loop induced deviation of $\triph$ from its tree-level value while the insert
presents the size of the corrections coming from the heavy
neutrino. With these mixing parameters, a heavy neutrino mass of
$m_{n_4}=2.7$~TeV corresponds to an effective coupling to the Higgs
equal to the one of the top quark while $m_{n_4}=7$~TeV leads to the
saturation of the tight perturbativity bound and $m_{n_4}=9$~TeV
saturates the width constraint. We observe that in the SM, the largest
positive correction is at $q_{H^*}^{}\simeq 500$~GeV, where the BSM
contribution decreases it to $-9\%$ at $m_{n_4}^{}=9$ TeV. The
largest negative correction comes at larger momentum where the
deviation from the SM increases with larger $m_{n_4}^{}$, reaching
$+30\%$ for $m_{n_4}^{}=9$ TeV at $q_{H^*}^{} = 2500$~GeV.

\begin{figure*}[!t]
   \centering
   \includegraphics[width=0.49\textwidth]{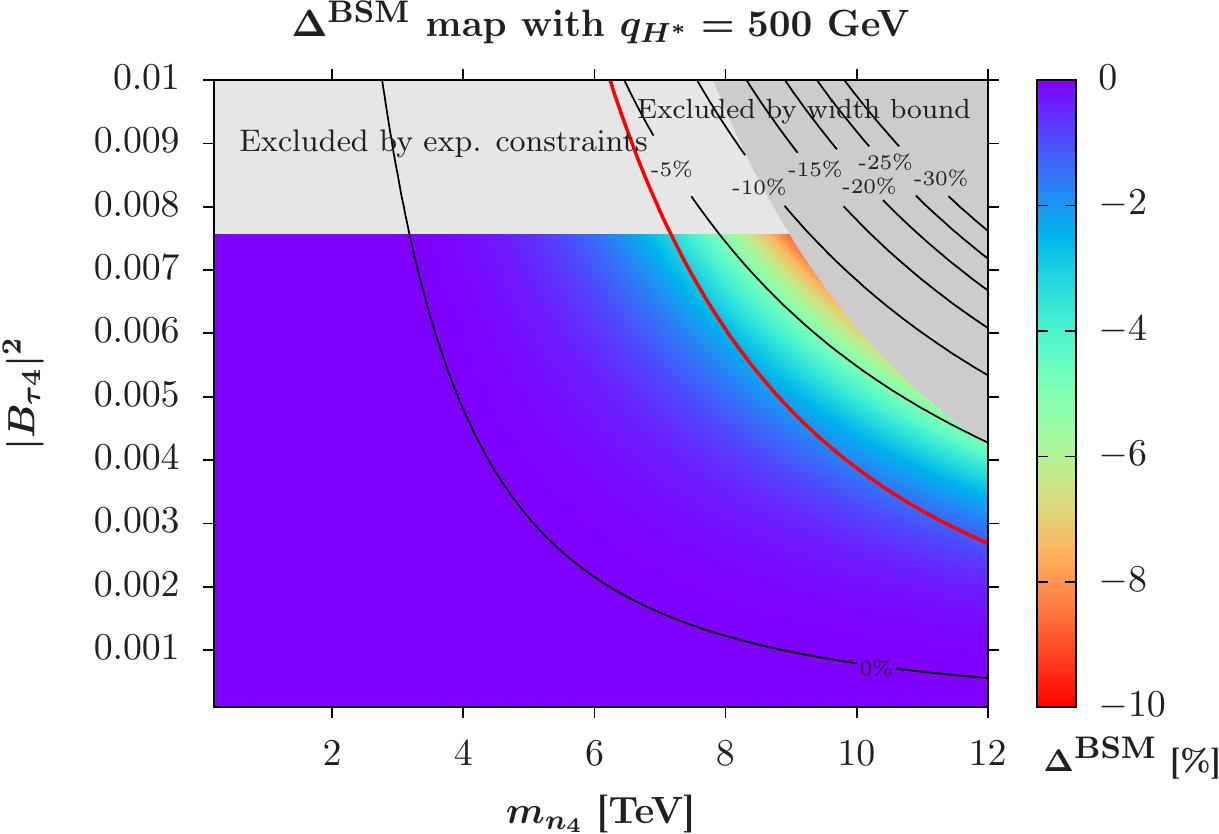}
   \includegraphics[width=0.49\textwidth]{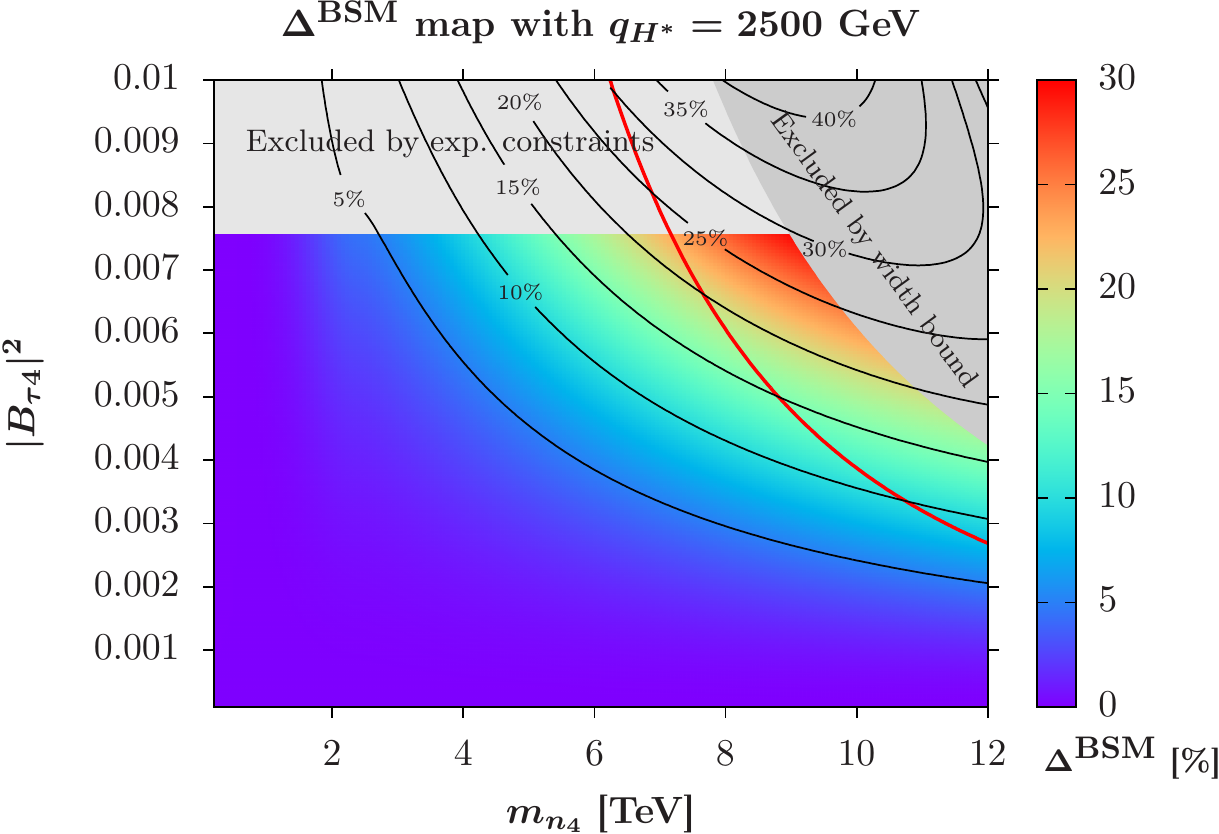}
   \caption{Contour maps of the neutrino corrections $\Delta_{}^{\rm
       BSM}$ to the triple Higgs coupling $\triph$ (in \%)
     as a function of the two neutrino parameters $|B_{\tau 4}^{}|_{}^2$ and
     $m_{n_4}^{}$ (in TeV), at a fixed off-shell Higgs momentum $q_{H^*}^{} =
     500$ GeV (left) and $q_{H^*}^{}=2500$ GeV (right). The other heavy
     neutrino mixing parameters are set to zero. The light grey area
     is excluded by the experimental constraints and the darker grey
     area is excluded from having $\Gamma_{n_4}^{}>0.6\,m_{n_4}^{}$ while
     the red line corresponds to the tight perturbativity bound.
     \label{fig:hhh-scan}}
 \end{figure*}

This behaviour leads us to chose $q_{H^*}^{}\simeq 500/2500$ GeV as
two most interesting off-shell momenta and to study the size of the
BSM corrections induced by the heavy neutrino as a function of its
mass and couplings. This is presented in Fig.~\ref{fig:hhh-scan} for
$q_{H^*}^{} = 500$~GeV (left) and $q_{H^*}^{} = 2.5$~TeV (right). The
largest effects are present in the high mixing / high heavy neutrino mass
region, reaching slightly less that $10\%$ negative deviation at
$q_{H^*}^{} = 500$~GeV (less that $-5\%$ with the tight perturbative
bound displayed in red) and around $+30\%$ increase at $q_{H^*}^{} =
2.5$~TeV (slightly less that $+25\%$ with the tight perturbative
bound). This is always below the HL-LHC sensitivity (35\%), but clearly
visible at the FCC-hh (5\%) and potentially visible at the ILC
(10\%).

\section{Inverse seesaw model}

In order to confirm the results obtained in the simplified 3+1 model,
we performed in our next study~\cite{Baglio:2016bop} the analysis of
the one-loop corrections to $\triph$ in a renormalisable, low-scale
seesaw model, namely the inverse seesaw
(ISS)~\cite{Mohapatra:1986bd,Mohapatra:1986aw,Bernabeu:1987gr}. 
We add to the SM Lagrangian 6 fermionic gauge singlets, 3 states
with positive lepton number $L=+1$ ($\nu_R^{}$) and 3 states with
negative lepton number $L=-1$  ($X$) with the following Yukawa couplings and mass terms,
\begin{align}
 \mathcal{L}_{\rm ISS} = & - Y_{\nu}^{i j} \overline{\ell_i^{}} \widetilde \Phi
                            \nu_{R j}^{} - M_R^{ij} \overline{\nu^C_{Ri}}
                            X_j^{} - \frac12 \mu_X^{ij} \overline{X^C_i} X_j^{}
                            + \mathrm{H.c.}\,,
\end{align}
leading after EWSB to 9 Majorana neutrinos $N_i^{}$. Thanks to
the two scale parameters $\mu_X^{}$ and $M_R^{}$ it is possible to
decouple the neutrino mass generation from the mixing between the
active neutrinos and the fermionic gauge singlets. The light neutrino
masses are suppressed by the small lepton-number breaking parameter
$\mu_X^{}$. It is then possible to have at the same time Yukawa couplings
$Y_{\nu}^{}$ of order 1 and $M_R^{} \sim 1$~TeV, which is within reach
of the LHC and low energy experiments.

The calculation of the one-loop corrections to $\triph$ in the ISS is
very similar to the calculation in the 3+1 model, but with Majorana
neutrinos instead of Dirac neutrinos. All formulae are available in
the appendix B of our article~\cite{Baglio:2016bop}. The set of 
constraints changes, though. The constraints from low-energy neutrino
data are implemented via the $\mu_X^{}$-parametrisation~\cite{Arganda:2014dta},
\begin{align}
\mu_X^{} = M_R^T m_D^{-1} U_{\rm PMNS}^* m_\nu^{} U_{\rm
  PMNS}^\dagger m_D^{T -1} M_R^{},
\end{align}
at the lowest order in the seesaw expansion parameter $m_D^{}
M_R^{-1}$. Terms beyond this order are also included in our analysis
and are given in the appendix A of our
article~\cite{Baglio:2016bop}. Charged lepton flavour violation
bounds are taken into account (see e.g. Ref.~\cite{TheMEG:2016wtm}) as
well as the global fit to EWPO and the lepton universality tests~\cite{Fernandez-Martinez:2016lgt}. 
Theoretical constraint on the heavy
neutrino widths $\Gamma_{N_i}^{} \leq 0.6 m_{N_i}^{}$ and the Yukawa
perturbativity constraint $|Y_\nu^{2}| < 1.5 \times 4 \pi$ are also
included.

\begin{figure*}[!t]
  \centering
  \includegraphics[width=0.49\textwidth]{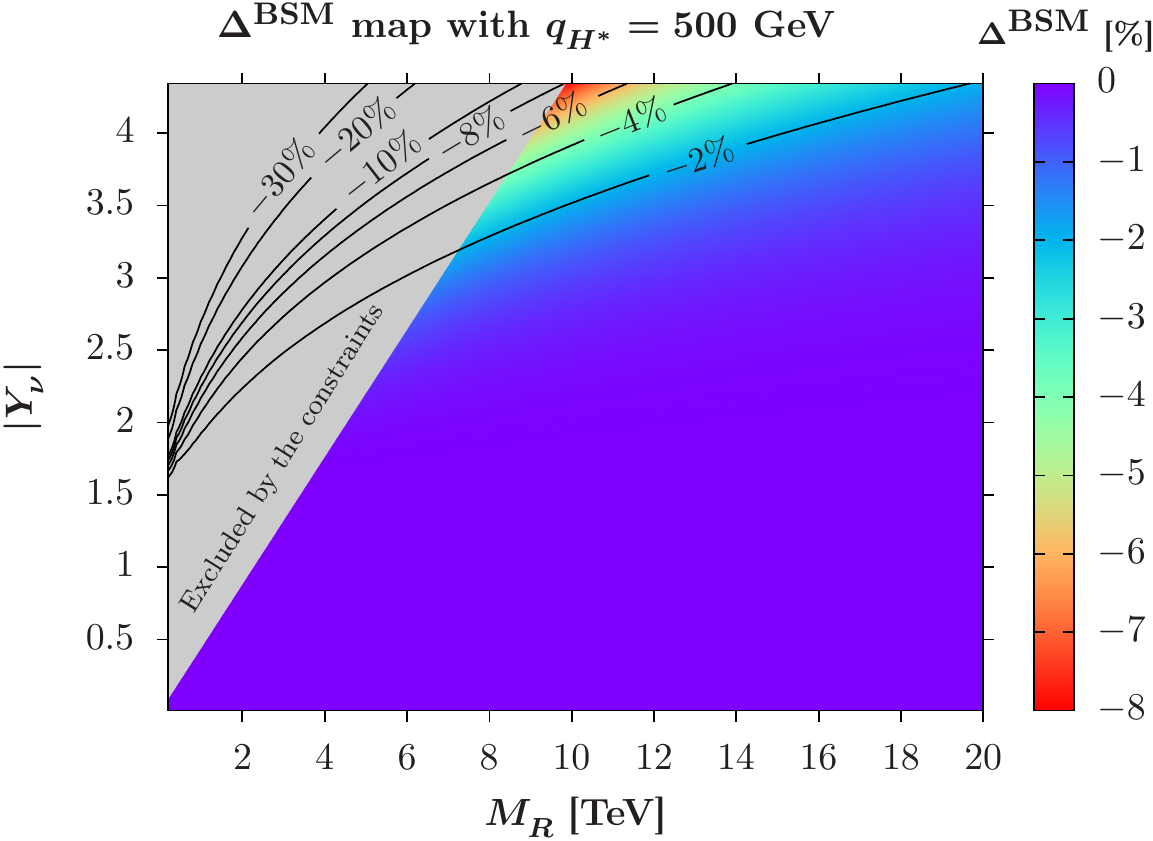}
  \includegraphics[width=0.49\textwidth]{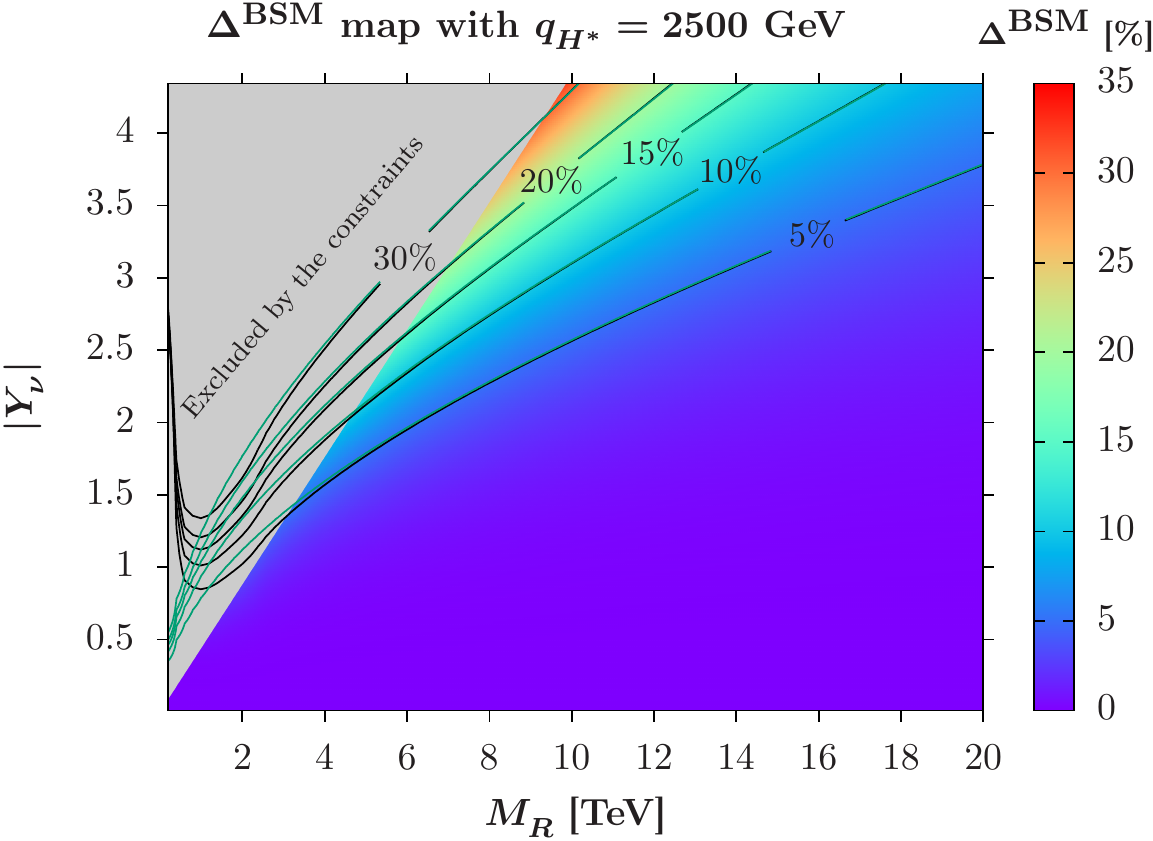}
  \caption{Contour maps of the heavy neutrino correction $\Delta^{\rm
      BSM}_{}$ to the triple Higgs coupling $\triph$ (in $\%$) as a
    function of the neutrino parameters $M_R^{}$ (in TeV) and
    $|Y_\nu^{}|$ in the $\mu_X^{}$--parametrisation, at a fixed
    off-shell Higgs momentum $q_{H^*}^{} = 500$~GeV and
    $q_{H^*}^{}=2.5$~TeV (right). We have used a diagonal Yukawa
    texture $Y_\nu^{}$ with parameter $|Y_\nu^{}|$ and a hierarchical
    heavy neutrino mass matrix, varying the seesaw scale $M_R^{}$. The
    grey area is excluded by the constraints on the model and the
    green lines on the right figure are the approximated contour lines
    using eq.(\protect\ref{eq:approx}), while the black lines
    correspond to the full calculation. 
    \label{fig:hierarchical}}
\end{figure*}

Representative numerical results of our study of the size of the BSM
corrections induced by the heavy neutrinos are displayed in
Fig.~\ref{fig:hierarchical}. We have used a diagonal Yukawa texture
$Y_\nu^{} = |Y_\nu^{}| I_3^{}$ and a hierarchical heavy neutrino mass
matrix with $M_R^{} = {\rm diag}(1.51 M_R^{}, 3.59 M_R^{}, M_R^{})$,
where $M_R^{}$ corresponds to the seesaw scale. This choice of textures for the two matrices leads to
the largest effects in the ISS, reaching the
$+30\%$ deviation for a large momentum $q_{H_{}^*}^{} = 2.5$~TeV and the 
$-8\%$ deviation for $q_{H_{}^*}^{} = 500$~GeV, for $M_R^{}$ around
9~TeV and large Yukawa couplings.
These results can be approximated by the following formula that is
used for the green lines in Fig.~\ref{fig:hierarchical},
 \begin{align}
 \Delta^{\rm BSM}_{\rm approx} = 0.51\frac{(1~\text{TeV})_{}^2}{M_R^2} \left( 8.45\, {\rm Tr}
   (Y_\nu^{} Y_\nu^\dagger Y_\nu^{} Y_\nu^\dagger) - 0.145\, {\rm Tr}
   (Y_\nu^{}   Y_\nu^\dagger Y_\nu^{} Y_\nu^\dagger Y_\nu^{}
   Y_\nu^\dagger)\right).
 \label{eq:approx}
 \end{align}
The larger number of heavy neutrinos compared to the simplified 3+1 model would naively induce larger corrections in the ISS.
However, the experimental constraints are stronger in this case, which leads to results similar to the simplified 3+1 analysis of the previous
section. This confirms our previous conclusion that the triple Higgs coupling $\triph$ is a
new, viable observable to constrain neutrino mass models.

\section{Conclusion}

The indubitable observation of neutrino oscillations requires the addition of BSM physics to 
generate neutrino masses and mixing and one of the simplest
and well-motivated ideas adds new right-handed sterile neutrinos to the SM,
leading to seesaw models. We have introduced the triple Higgs coupling
as a new observable to constraint these neutrino mass models and we have
found, first in a simplified 3+1 model and then in the inverse seesaw,
that the one-loop effects induced by these heavy neutrinos can be
as large as $30\%$ with respect to the SM one-loop prediction. This is
measurable at future colliders and it provides a new,
complementary probe in the $\mathcal{O}(10)$~TeV range of the heavy
neutrino masses. We stress that these effects are generic and would be
expected in any model containing TeV scale fermions with large Higgs
couplings, such as those using the neutrino portal.  

\acknowledgments{
J.~B. acknowledges the support from the Institutional Strategy of the
University of T\"ubingen (DFG, ZUK 63), from the DFG Grant JA 1954/1.,
from the Kepler Center of the University of T\"ubingen, as well as
from his Durham Senior Research Fellowship COFUNDed between Durham
University and the European Union under grant agreement number
609412. C.~W acknowledges the support from the European Research
Council under the European Union’s Seventh Framework Programme
(FP/2007-2013) / ERC Grant NuMass agreement n. [617143] and partial
support from the European Union's Horizon 2020 research and innovation
programme under the Marie Sk\l{}odowska-Curie grant agreements
No.~690575 and No.~674896.


\begin{thebibliography}{99}
\bibitem{Fukuda:1998mi}
Super-Kamiokande collaboration, Y.~Fukuda et~al., \
  Phys.\ Rev.\ Lett.\ {\bf 81} (1998) 1562--1567.

\bibitem{Higgs:1964ia}
  P.~W.~Higgs, Phys.\ Lett.\  {\bf 12} (1964) 132.

\bibitem{Englert:1964et}
  F.~Englert and R.~Brout, Phys.\ Rev.\ Lett.\  {\bf 13} (1964) 321.

\bibitem{Higgs:1964pj}
  P.~W.~Higgs, Phys.\ Rev.\ Lett.\  {\bf 13} (1964) 508.

\bibitem{Guralnik:1964eu}
  G.~S.~Guralnik, C.~R.~Hagen and T.~W.~B.~Kibble,
  Phys.\ Rev.\ Lett.\  {\bf 13} (1964) 585.

\bibitem{Higgs:1966ev}
  P.~W.~Higgs, Phys.\ Rev.\  {\bf 145} (1966) 1156.

\bibitem{Mohapatra:1986bd}
  R.~N.~Mohapatra and J.~W.~F.~Valle,
  Phys.\ Rev.\ D {\bf 34} (1986) 1642.

\bibitem{Mohapatra:1986aw}
  R.~N.~Mohapatra,
  Phys.\ Rev.\ Lett.\  {\bf 56} (1986) 561.

\bibitem{Bernabeu:1987gr}
  J.~Bernabeu, A.~Santamaria, J.~Vidal, A.~Mendez and J.~W.~F.~Valle,
  Phys.\ Lett.\ B {\bf 187} (1987) 303.

\bibitem{Baglio:2016ijw}
  J.~Baglio and C.~Weiland,
  Phys.\ Rev.\ D {\bf 94} (2016) 013002.

\bibitem{Baglio:2016bop}
  J.~Baglio and C.~Weiland, JHEP {\bf 1704} (2017) 038.

\bibitem{CMS:2015nat}
  CMS Collaboration [CMS Collaboration],
  CMS-PAS-FTR-15-002.
  
\bibitem{Campana:2016cqm}
  P.~Campana, M.~Klute and P.~Wells,
  Ann.\ Rev.\ Nucl.\ Part.\ Sci.\  {\bf 66} (2016) 273.
  
\bibitem{He:2015spf}
  H.~J.~He, J.~Ren and W.~Yao,
  Phys.\ Rev.\ D {\bf 93} (2016)  015003.

\bibitem{Fujii:2015jha}
  K.~Fujii {\it et al.},
  arXiv:1506.05992 [hep-ex].

\bibitem{delAguila:2008pw}
  F.~del Aguila, J.~de Blas and M.~Perez-Victoria,
  Phys.\ Rev.\ D {\bf 78} (2008) 013010.
 
\bibitem{deBlas:2013gla}
  J.~de Blas,
  EPJ Web Conf.\  {\bf 60} (2013) 19008.

\bibitem{Gonzalez-Garcia:2014bfa}
  M.~C.~Gonzalez-Garcia, M.~Maltoni and T.~Schwetz,
  JHEP {\bf 1411} (2014) 052.

\bibitem{Patrignani:2016xqp}
  C.~Patrignani {\it et al.} [Particle Data Group],
  Chin.\ Phys.\ C {\bf 40} (2016) 100001.

\bibitem{Arganda:2014dta}
  E.~Arganda, M.~J.~Herrero, X.~Marcano and C.~Weiland,
  Phys.\ Rev.\ D {\bf 91} (2015) 015001.
  
\bibitem{TheMEG:2016wtm}
  A.~M.~Baldini {\it et al.} [MEG Collaboration],
  Eur.\ Phys.\ J.\ C {\bf 76} (2016) no.8,  434.

\bibitem{Fernandez-Martinez:2016lgt}
  E.~Fernandez-Martinez, J.~Hernandez-Garcia and J.~Lopez-Pavon,
  JHEP {\bf 1608} (2016) 033.

\end{thebibliography}
\end{document}